\documentclass[nolinenumbers]{aastex631}

\usepackage[utf8]{inputenc}
\usepackage{amsmath}
\usepackage{amssymb}
\usepackage{graphicx}
\usepackage{bm}
\usepackage{float}
\usepackage{hyperref}
\hypersetup{colorlinks=true, linkcolor=blue,filecolor=blue,urlcolor=blue,citecolor=blue}

\begin{document}

\title{The infrared colors of 51 Eridani b: \protect\\ micrometereoid dust or chemical disequilibrium?}

\author{Alexander Madurowicz}
\affiliation{Kavli Institute for Particle Astrophysics and Cosmology, Stanford University}
\author{Sagnick Mukherjee}
\affiliation{Department of Astronomy and Astrophysics, University of California, Santa Cruz}
\author{Natasha Batalha}
\affiliation{NASA Ames Research Center}
\author{Bruce Macintosh}
\affiliation{Kavli Institute for Particle Astrophysics and Cosmology, Stanford University}
\affiliation{Department of Astronomy and Astrophysics, University of California, Santa Cruz}
\author{Mark Marley}
\affiliation{Lunar and Planetary Laboratory, The University of Arizona}
\author{Theodora Karalidi}
\affiliation{Department of Physics, University of Central Florida}

\begin{abstract}
    We reanalyze near-infrared spectra of the young extrasolar giant planet 51 Eridani b which was originally presented in \citep{macintosh2015} and \citep{Rajan_2017} using modern atmospheric models which include a self-consistent treatment of disequilibrium chemistry due to turbulent vertical mixing. In addition, we investigate the possibility that significant opacity from micrometeors or other impactors in the planet's atmosphere may be responsible for shaping the observed spectral energy distribution (SED). We find that disequilibrium chemistry is useful for describing the mid-infrared colors of the planet's spectra, especially in regards to photometric data at M band around 4.5 $\mu$m which is the result of super-equilibrium abundances of carbon monoxide, while the micrometeors are unlikely to play a pivotal role in shaping the SED. The best-fitting, micrometeroid-dust-free, disequilibrium chemistry, patchy cloud model has the following parameters: effective temperature $T_\textrm{eff} = 681$ K with clouds (or without clouds, i.e. the grid temperature $T_\textrm{grid}$ = 900 K), surface gravity $g$ = 1000 m/s$^2$, sedimentation efficiency $f_\textrm{sed}$ = 10, vertical eddy diffusion coefficient $K_\textrm{zz}$ = 10$^3$ cm$^2$/s, cloud hole fraction $f_\textrm{hole}$ = 0.2, and planet radius $R_\textrm{planet}$ = 1.0 R$_\textrm{Jup}$.
\end{abstract}

\section{Introduction}
Direct imaging \citep{pueyo2018,Bowler2016} is a powerful tool for detection and characterization of extrasolar planets. Astrometric measurements constrain planetary orbital elements \citep{Konopacky_2016} and dynamical stability of multi-body systems \citep{Wang_2018} constrains their masses. Spectroscopic measurements at low resolution reveal molecular abundances, atmospheric parameters, and cloud properties \citep{Ingraham_2014}. At higher resolution, spectroscopy can even measure planetary radial velocities and spin rotation rates \citep{Wang_2021,Snellen2014}.

However, observations on their own are insufficient and theoretical models which attempt to reproduce the data are necessary to provide physical context and enable interpretations of the empirical data. These models provide the necessary framework for performing a robust inference of planetary parameters, but the fidelity of those inferences is then ultimately rooted in the accuracy of the assumptions underlying the computational models used in making them. Existing observations of the exoplanet 51 Eridani b \citep{Rajan_2017,macintosh2015} demonstrate this notion. By assuming the atmospheric chemistry is in equilibrium, estimates of the abundance of carbon monoxide are too low resulting in a spectrum which is too bright at mid-infrared wavelengths. Investigations of isolated brown dwarfs \citep{Miles_2020, griffith2000, Zahnle_2014} as well as transiting hot Jupiters \citep{claire2021} have demonstrated that disequilibrium chemistry \citep{mukherjee2022, Saumon1996, Noll_1997, marley_robinson_2015}, specifically the presence carbon monoxide produced by atmospheric quenching,  is critical for reproducing the spectral colors of substellar objects.

Additionally, just as the radiatively accessible upper layers of the planetary atmosphere may be modified by turbulent dynamics dredging up different molecules from the hotter and deeper layers, the upper boundary condition may influence the appearance of the atmosphere as well. Extreme events such as the comet Shoemaker-Levy 9's \citep{hammel1995} impact with Jupiter provide a spectacular example. Observations suggest that as a result of the impact the atmospheric thermal profile and composition are substantially altered for weeks after the impact \citep{Lellouch1995}.  Extremely young protoplanets appear bright in H-$\alpha$ in the ultraviolet due to ongoing accretion in the protoplanetary disk \citep{Zhou_2021}. The possibility that interplanetary or circumstellar dust captured by exoplanets could modify their spectra has recently been investigated \citep{Arras_2022} in the context of transiting planets. Since systems which host debris disks are much more likely to host planets than those without \citep{Meshkat_2017, Marshall2014}, considering the interactions between planets and disk material is important. The dustiness of an extrasolar system can be quantified by the ratio of infrared emission to stellar luminosity $L_\textrm{IR}/L_{*}$. For 51 Eridani in particular $L_\textrm{IR}/L_{*} = 2.3 \times 10^{-6}$ \citep{Riviere-Marichalar2014}, while for the solar system this value is an order of magnitude smaller \citep{Wyatt2008}. But the system with the most prominent debris disks can have $L_\textrm{IR}/L_{*} \sim 10^{-2}$ \citep{Esposito2020} and this generically evolves over time with a power law index of -2 \citep{Spangler2001}. Previous studies of directly imaged substellar objects \citep{Ward_Duong_2020, Cushing_2006, Marocco2015, Hiranaka2016, Burningham2021}
have invoked sub-micron sized dust particles as a potential mechanism for reddening spectra beyond what typical model grids can reproduce. 

Ultimately, a complete theory of planetary atmospheres synthesizes first principles theory with observations \citep{Zhang_2020} to better understand their complex nature. In this paper, we investigate two possible mechanisms which modify the spectral colors of the exoplanet 51 Eridani b, micrometeroid dust rain and disequilibrium chemistry in the atmosphere. We present our results as a model comparison and parameter inference. Section \ref{methods} discusses the methodology behind our analysis. We briefly discuss the observational data and software tools used to calculate cloud condensate profiles and radiative transfer. The most important results in this section showcase the distinct effects on the final spectra which result from changing the atmospheric chemistry or including micrometeroid dust in the atmosphere. Section \ref{results} explores a grid of 374,673,600 1D radiative-convective atmospheric models, demonstrating best fit model spectra from each class of models, as well as triangle plots of posterior parameter inferences for each class over the entire parameter grid. Section \ref{discussion} concludes the paper with a short discussion on potential further improvements to the model fidelity. The majority of the mathematical description of the model of micrometeroid dust is enumerated after the discussion in an Appendix.

\section{Methods}\label{methods}

The observations of 51 Eridani b used in this paper were originally presented and published in \citep{Rajan_2017}. The observations include spectroscopy in near infrared bands J (1.13–1.35 $\mu$m), H (1.50–1.80 $\mu$m), K1 (1.90–2.19 $\mu$m), and K2 (2.10–2.40 $\mu$m) taken with the Gemini Planet Imager (GPI) on the Gemini South telescope, as well as photometric points at moderate infrared bands L$_P$ (3.43-4.13 $\mu$m) and M$_S$ (4.55-4.79 $\mu$m) taken with NIRC2 on the Keck Telescope. The GPI data are processed according to standard data reduction procedures laid out in \citep{perrin2014} including dark current subtraction, removal of bad pixels, corrections for instrument flexure; as well as extraction, interpolation, distortion correction, and alignment of microspectra for producing IFS cubes \citep{maire2014}. More details on the observational strategies and data reduction for the Keck observations can be found in the original paper.

To model the spectra of 51 Eridani b we use a suite of existing software with a rich heritage in modeling giant planets and brown dwarfs. \texttt{PICASO 3.0} \citep{mukherjee2022} is used for determining the atmospheric thermal structure using a self-consistent treatment of disequilibrium chemistry. \texttt{Virga} \citep{virga2020} is used for computing cloud condensate profiles, essentially vertical number densities and particle size distributions of each relevant chemical species according to eddy-diffusion and sedimentation equilibrium \citep{Ackerman_2001}. In particular, the model includes condensate clouds of the following molecules and elements: Al$_2$O$_3$, Cr, Fe, KCl, Mg$_2$SiO$_4$, MgSiO$_3$, MnS, Na$_2$S, TiO$_2$, and ZnS. \texttt{Virga} relies on extensive published empirical data \citep{querry1987,huffman1967,1984stashchuk,montaner1979,Khachai_2009,scott1996,leksina1967,KOIKE1995,Martonchik84,henning1999,jager2003} for determining the optical properties of relevant condensates such as the complex index of refraction. Standard mie theory \citep{dave1968subroutines,Sumlin2018} is used to translate these values into optical scattering properties as a function of wavelength $\lambda$ and pressure altitude $P$ including single scattering albedo $w_0 (P, \lambda)$, optical depth per layer $\textrm{OPD}(P, \lambda)$, and the scattering asymmetry parameter $g_0(P, \lambda)$ which are the fundamental inputs to \texttt{PICASO}. These values are plotted in the left panels of Figure (\ref{fig:cloud}). Lastly, \texttt{PICASO} \citep{Batalha_2019} is used for performing radiative transfer calculations to obtain the final spectra. Additionally, the model includes a patchy cloud framework based on \citep{Marley_2010} governed by the cloud hole fraction parameter $f_\textrm{hole}$. It is important to note that the addition of patchy clouds and micrometeroid opacity into the radiative transfer is not entirely self-consistent, as the cloud radiative feedback on the thermal structure of the atmosphere is ignored, and each scattering event is effectively dissipating energy which would otherwise reheat the atmosphere. This results in two different measurements of the effective temperature of the model: $T_\textrm{grid}$ which is the effective temperature without the clouds, using the self-consistent disequilibrium framework of \citep{mukherjee2022}, and the final $T_\textrm{eff}$ which is lower due to the effect of the clouds. The discrepancy between these two effective temperatures indicates the relative importance of treating cloud radiative feedback completely self-consistently, but is outside the scope of this work.

After the radiative transfer has completed, and in order to compare models and infer optimal parameters, we use the parameterized model of spectral covariance laid out in the appendix of \citep{Rajan_2017} and originally derived by \citep{Greco_2016}. In the processing of IFS data into spectral cubes, interpolation of pixel values into wavelength bins results in data which are not truly independent measurements. It is therefore critically important to include the covariance of spectral data points taken with integral field spectrographs to avoid biasing the resulting parameter inferences. The model includes three distinct terms, one image location and wavelength dependent term which attempts to account for speckle noise, one wavelength dependent term to account for the interpolation, and an uncorrelated term to account for read noise. The covariance model is estimated due to its high dimensionality with an MCMC based sampling procedure on the PSF-subtracted data. Finally, the quality of the model fit is estimated with a $\chi^2$ analysis over all of the data points, where the covariance $C$ is used for the spectra and the simple uncertainties of the data points $\sigma$ are used for the photometric points. Implicitly both $C$ and $\sigma$ are unique for each specific spectroscopic band or photometric point in the sum. 
\begin{equation}
    \chi^2 = \sum_{J,H,K_1,K_2} v^T C^{-1} v + \sum_{L,M} \Big(\frac{v}{\sigma}\Big)^2
    \label{eq:chi2}
\end{equation}
Here $v$ is a vector which represents the difference between the modelled and observed flux at each wavelength
\begin{equation}
    v = \Big(\frac{R_\textrm{planet}}{d_*}\Big)^2 F_{\lambda,\textrm{model}} - F_{\lambda,\textrm{observation}},
\end{equation}
where $F_{\lambda,\textrm{model}}$ is the computed top-of-the-atmosphere flux, which is rescaled by the geometry of planet and system using the inverse square law for radiation to correspond the observed flux $F_{\lambda,\textrm{observed}}$. Furthermore, the reduced chi-square $\chi^2_\nu$, or chi-square per degree of freedom is used to compare models with and without micrometeroid dust added, which account for two additional free parameters in the model.

\subsection{Equilibrium versus Disequilibrium chemistry}
Unambiguous detections of carbon monoxide in late L-Type to T-Type brown dwarfs with AKARI \citep{Sorahana_2012}, as well as detections of carbon monoxide in Gliese 229 B \citep{Noll_1997,Oppenheimer_1998,Saumon_2000}, Gliese 570 D and 2MASS J09373487+2931409 \citep{Geballe_2009}, VHS 1256-1257 b \citep{miles2022}, demonstrate that M-band absorption due to super-equilibrium abundances of carbon monoxide are not only commonplace in substellar atmospheres, but necessary to properly model their spectra. Similar processes have been speculated to influence the colors of young, massive directly imaged giant planets \citep{marley_robinson_2015}, but existing detection of carbon monoxide in exoplanets \citep{brogi2014,konopacky2013,Barman_2011} do not produce precise enough constraints to warrant an exploration of disequilibrium abundances.

\begin{figure}[htbp]
    \centering
    \includegraphics[width=\textwidth]{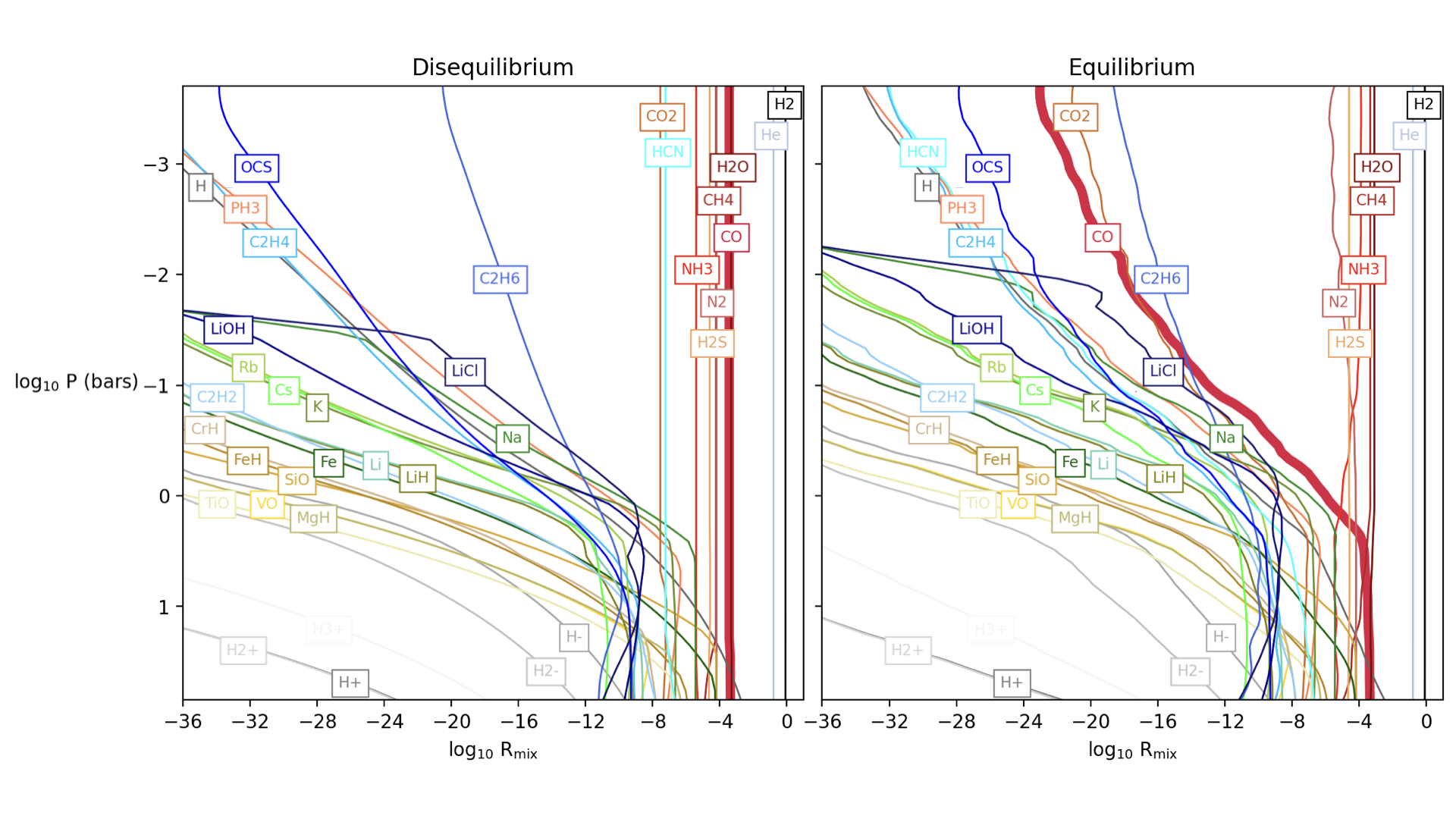}
    \caption{Molecular volume mixing ratios comparing the equilibrium and disequilibrium abundances as a function of pressure altitude. The significant presence of CO in the disequilibrium model is highlighted with a thick curve. Molecules are roughly grouped into color families and labels are constrained to the same y-coordinate in both panels. Additional species noticeably affected by the disequilibrium chemistry include N$_2$, NH$_3$, CO$_2$, and HCN.}
    \label{fig:diseq}
\end{figure}

Past models commonly assumed that molecules of various chemical species in the planet's atmosphere are in equilibrium \citep{fortney2015, marley2017}, but other models incorporate disequilibrium abundances \citep{Karalidi_2021, mukherjee2022, Hubeny_2007, Phillips2020}. We provide a comparison of molecular abundances for the two different assumptions in Figure (\ref{fig:diseq}).

The disequilibrium abundances are largely based on the ``quench approximation'' \citep{marley_robinson_2015,mukherjee2022}, where abundances follow equilibrium chemistry when the mixing timescale $t_\textrm{mix} \approx \frac{H^2}{K_\textrm{zz}}$ is much larger than the chemical timescale $t_\textrm{mix} >> t_\textrm{chem}$. Here $H$ is the atmospheric scale height and $K_\textrm{zz}$ the vertical eddy diffusion coefficient, so that higher $K_\textrm{zz}$ implies greater mixing and thus shorter timescale for mixing. This condition is reached deep in the atmosphere, at pressures higher than the ``quench pressure.'' At higher altitudes or equivalently lower pressures, the chemical timescale is long compared to mixing, and the abundances are quenched to a constant value. The chemical timescales are estimated from one dimensional chemical kinetics models \citep{Zahnle_2014}, see section 2.1.5 in \citep{mukherjee2022} and 5.3 in \citep{marley_robinson_2015} for more details.

\begin{figure}[htbp]
    \centering
    \includegraphics[width=\textwidth]{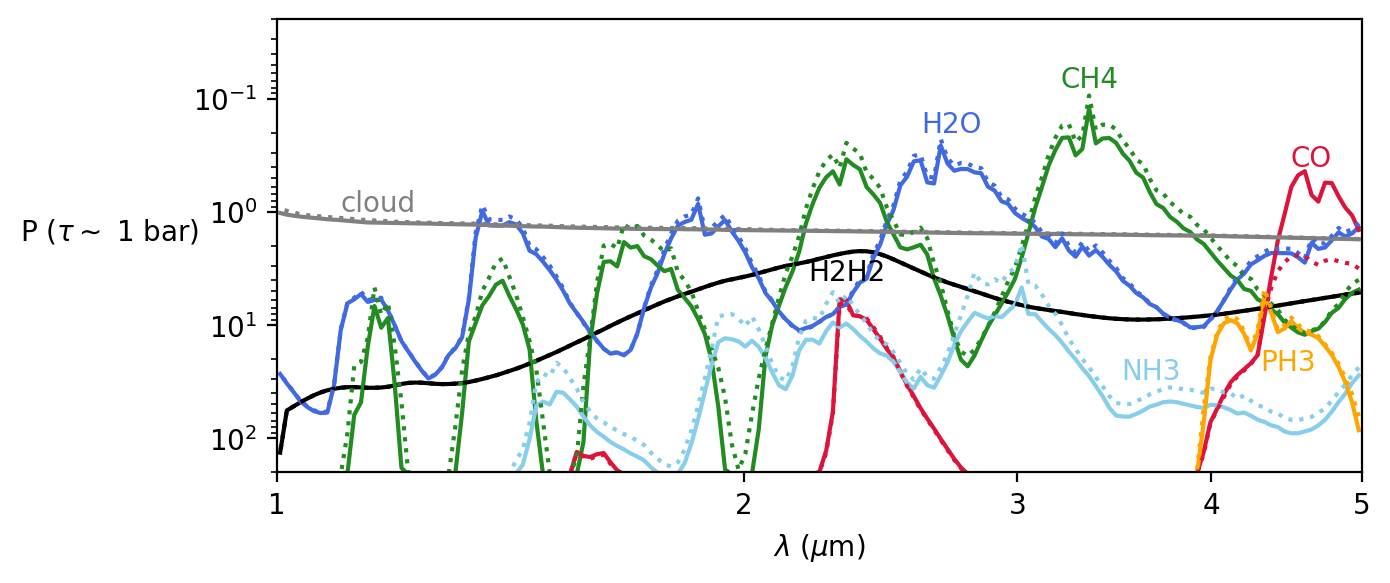}
    \caption{Optical contribution of major gaseous absorbers in the disequilibrium (solid) and equilibrium (dotted) chemistry models. Curves correspond to pressures where the approximate cumulative optical depth per species is order unity. The high altitude CO in the disequilibrium model is evidently responsible for additional absorption in the range of 4-5 $\mu$m. This feature is generically visible in all of the disequilibrium models, but exactly how much varies with atmospheric parameters. This specific realization uses grid temperature $T_\textrm{grid} = 800$ K, surface gravity $g = 316$ m/s, sedimentation efficiency $f_\textrm{sed} = 10$, and vertical eddy diffusion coefficient $\log \frac{K\textrm{zz}}{\textrm{cm}^2/\textrm{s}} = 5$.}
    \label{fig:tau1}
\end{figure}

The effect which a large quantity of high altitude carbon monoxide has on the resulting spectra of a giant extrasolar planet is demonstrated in Figure (\ref{fig:tau1}). Using the \texttt{PICASO} built-in function \texttt{picaso.justdoit.get\_contribution}, we estimate the pressure altitude as a function of wavelength where the optical depth per species is of order unity, and filter out the species which are optically irrelevant or nearly so. This figure is a useful diagnostic to determine which species have the most outstanding influence on the resulting spectral energy distribution, as well as the relative contributions for absorption bands across various wavelengths.

The primary observation to take away from this figure is how the disequilibrium abundance of CO at high altitude induces significant absorption in the M band around 4.5 $\mu$m compared to the equilibrium chemistry model. The influence of other relevant species in the atmosphere is apparent as well, including prominent absorption features from methane and water vapor, continuum absorption from collision-induced absorption of diatomic hydrogen gas, and the opacity of the condensate clouds. The carbon monoxide feature relevant for the other detections in extrasolar planets around 2.3 $\mu$m is visible as well, although it is subdominant in this particular case due to the cooler thermal profile of 51 Eridani b. Additionally, subtle shifts in the absorption for important molecules such as methane and water vapor across all wavelengths lead to relatively large changes in the thermal structure of the atmosphere, which has a significant impact on the resulting SEDs.

\subsection{Micrometeroid Dust}

In addition to absorption from gaseous species and condensate clouds in the atmosphere, we consider the possibility that micrometeroids falling into the atmosphere from the circumplanetary environment could provide an additional source of opacity relevant for shaping the spectra of a planet. The details of our micrometeroid model are enumerated carefully in the Appendix. The model broadly corresponds to a bounded power-law size distribution of purely scattering, non-absorbing SiO$_2$ spheres inbound with a time-constant and surface-area-uniform number density flux. The silicate spheres fall through the atmosphere at a velocity governed by their terminal speed neglecting buoyancy, additional perturbations such as frictional ablation and heating, additional chemistry and radiative perturbations, spatial and temporal non-uniformity, non-sphericity and fragmentation of the rocky grains, uplift of the micrometeroids due to vertical winds, among other complex process which may shape the atmosphere and infalling rocky material. The model could be improved, but also should be sufficient as a preliminary investigation. 
\begin{figure}[htbp]
    \centering
    \includegraphics[width=.9\textwidth]{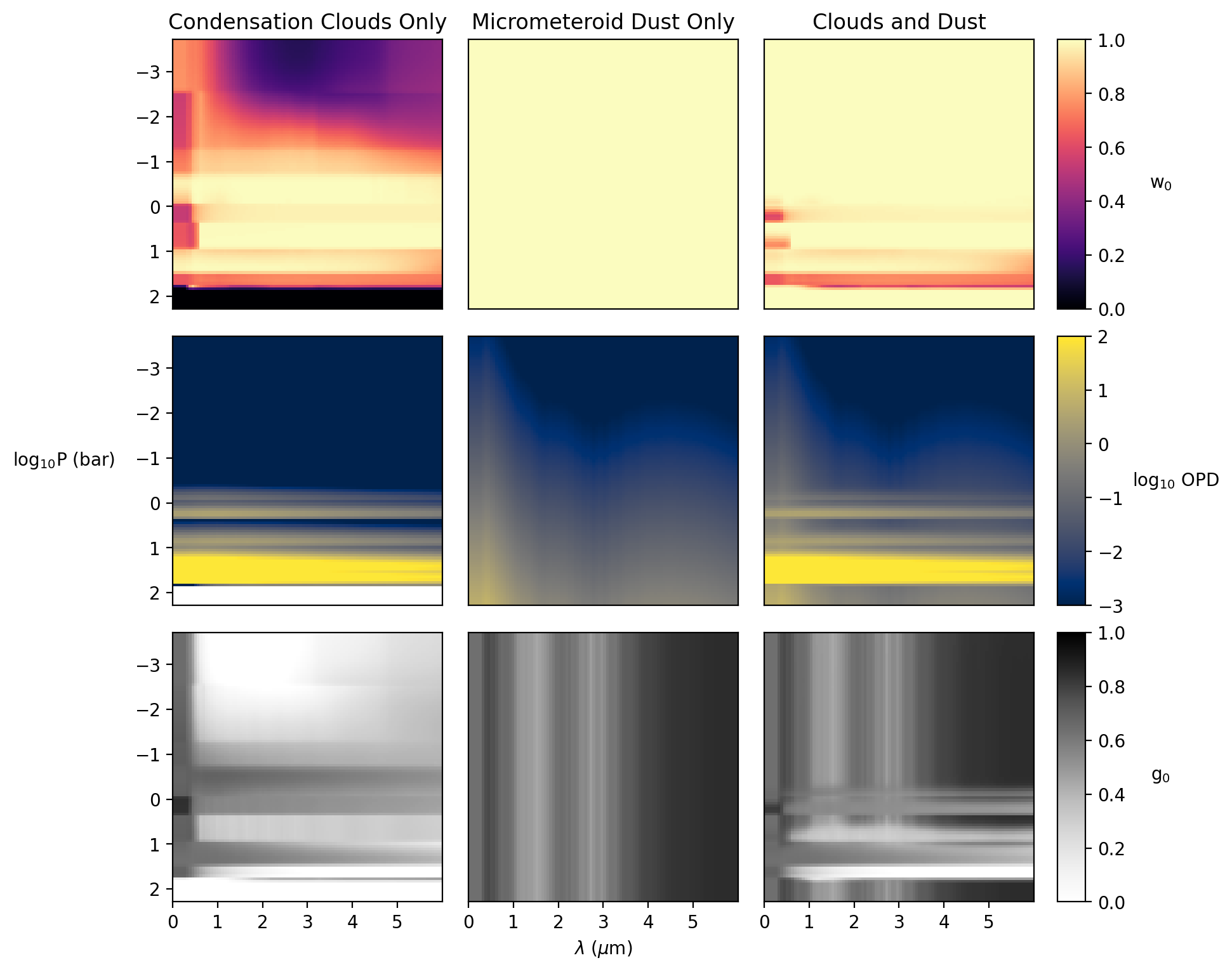}
    \caption{Single scattering albedo $w_0$, optical depth per layer OPD, and asymmetry parameter $g_0$, for the condensation cloud model computed with \texttt{Virga} as well as for the micrometeroid dust model. The combination of the two is the optical-depth weighted average for the scattering parameters or the sum of the optical depths themselves. This particular simulation uses grid temperature $T_\textrm{grid} = 800$ K, surface gravity $g = 316$ m/s, vertical eddy diffusion coefficient $K_{zz} = 10^{5}$ cm$^2$/s, sedimentation efficiency $f_\textrm{sed} = 10$, dust power law index $\alpha = -3.5$, and dust power law coefficient $\log N_0 = 14$.}
    \label{fig:cloud}
\end{figure}
The output of \texttt{Virga} is the three cloud model parameters (single scattering albedo $w_0$, optical depth per layer $\textrm{OPD}$, and the scattering asymmetry parameter $g_0$) as a function of wavelength and pressure altitude. In the Appendix we detail our calculations for computing these parameters for the additional micrometeroid dust in the atmosphere. In order to combine the cloud and dust models, we simply sum the optical depths per layer at every layer
\begin{equation}
    \textrm{OPD}_\textrm{combined}(P, \lambda) = \textrm{OPD}_\textrm{cloud}(P,\lambda) + \textrm{OPD}_\textrm{dust}(P, \lambda),
\end{equation}
and compute the optical-depth weighted average of the asymmetry parameter and single scattering albedo
\begin{equation}
    w_{0,\textrm{combined}}(P,\lambda) = \frac{w_{0,\textrm{cloud}}(P,\lambda) \textrm{OPD}_\textrm{cloud}(P,\lambda) + w_{0,\textrm{dust}}(P,\lambda) \textrm{OPD}_\textrm{dust}(P,\lambda)}{\textrm{OPD}_\textrm{cloud}(P, \lambda) + \textrm{OPD}_\textrm{dust}(P, \lambda)} 
\end{equation}
\begin{equation}
    g_{0,\textrm{combined}}(P,\lambda) = \frac{g_{0,\textrm{cloud}}(P,\lambda) \textrm{OPD}_\textrm{cloud}(P,\lambda) + g_{0,\textrm{dust}}(P,\lambda) \textrm{OPD}_\textrm{dust}(P,\lambda)}{\textrm{OPD}_\textrm{cloud}(P, \lambda) + \textrm{OPD}_\textrm{dust}(P, \lambda)}.
\end{equation}
Figure (\ref{fig:cloud}) demonstrates the impact of including micrometeroid dust on the cloud properties as a function of pressure altitude and wavelength. The impact of the micrometeors is most apparent as a high altitude source of wavelength dependent opacity above the condensate cloud decks.

\begin{figure}[htbp]
    \centering
    \includegraphics[width=\textwidth]{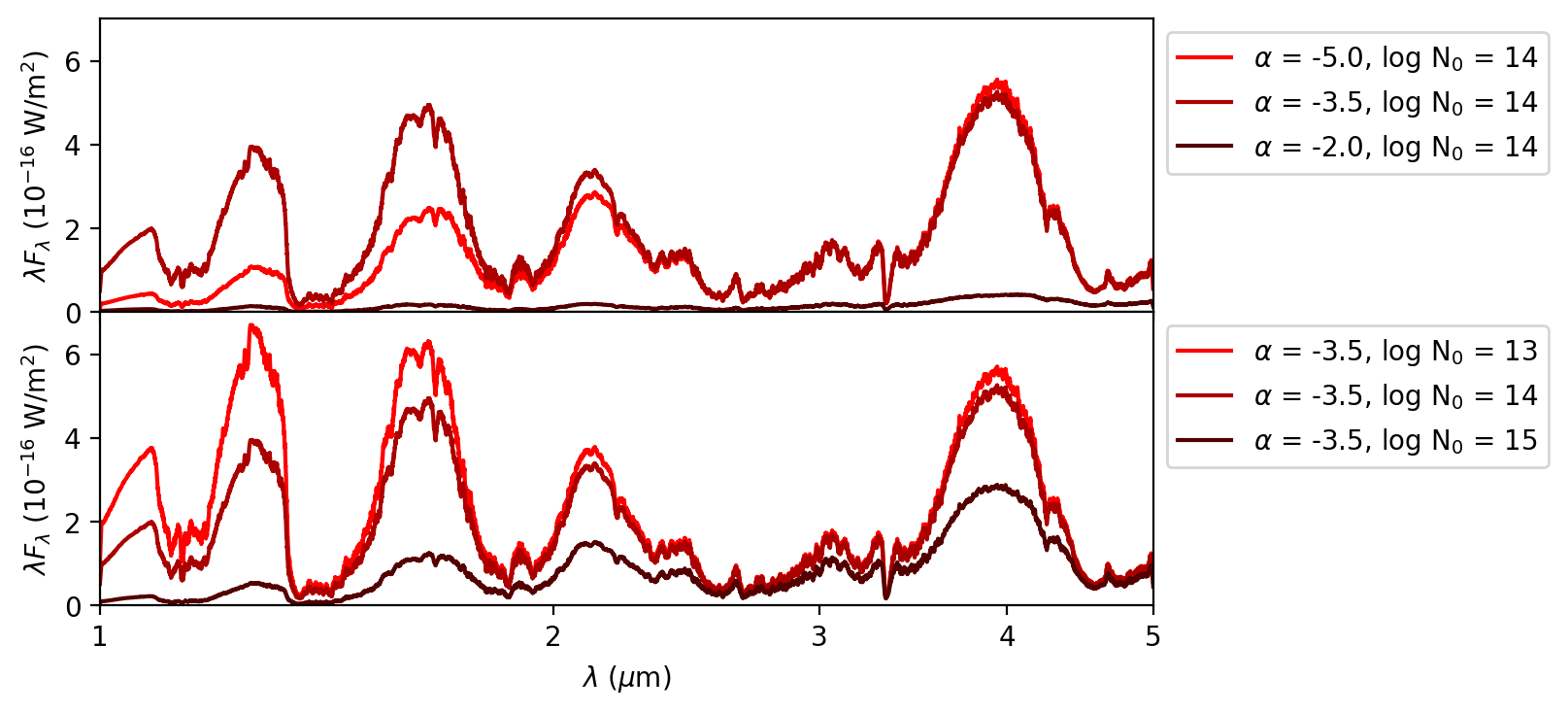}
    \caption{Influence of micrometeroid dust model parameters on the resulting spectral energy distribution. This particular simulation uses grid temperature $T_\textrm{grid} = 1000$ K, surface gravity $g = 316$ m/s, vertical eddy diffusion coeffficient $K_{zz} = 10^{10}$ cm$^2$/s, sedimentation efficiency $f_\textrm{sed} = 3$, and cloud hole fraction $f_\textrm{hole} = 0$, while the dust power law index $\alpha$ and coefficient $N_0$ vary depending on the line color in the legend. In the top panel $\alpha$ is varied, showing how steep power laws have the majority of optical power in sub-micron sized particles that scatter light at shorter wavelengths, while the flat power laws are dominated by sub-mm sized grains which scatter strongly at all relevant wavelengths. In the bottom panel the effect of increasing $N_0$ at fixed $\alpha$ is demonstrated. More infalling dust generically implies greater scattering, although the parameters chosen here represent potentially unphysical and extreme values which are necessary increase the significance of the effect and to make the resulting change in the spectra visible by eye.}
    \label{fig:dust}
\end{figure}

The impact of including micrometeroid dust in the cloud model adds two extra degrees of freedom to control the reddening and total brightness of the spectral energy distribution. Figure (\ref{fig:dust}) presents a visual demonstration of the influence of tuning the dust model parameters on the resulting spectra.
Changing $\alpha$, the dust power law spectral index, shifts the micrometeroid distribution to contain more or less millimeter-, micron-, or nanometer-sized grains falling into the planet's atmosphere. In general, grains much larger than the relevant wavelengths scatter strongly across all the wavelengths, while grain sizes smaller than the wavelength are much less efficient at scattering, with a noticeable influence at shorter wavelengths. However, changing $N_0$, the dust power law proportionality constant, alters the total number density of the particle flux into the atmosphere for any particular index $\alpha$. Larger $N_0$ implies greater quantities of dust and therefore implies greater scattering of emitted radiation, and thus dimming of the entire spectra. But a significant influence on the spectra is only noticeable for extremely large mass accretion fluxes, see Figure (\ref{fig:mass_infall}) in the Appendix for greater specificity, but for $\alpha = -3.5$ and $\log N_0 = 14$, the mass rate is of order 10$^{-10}$ M$_\odot$/yr, which is comparable to the accretion rate of gas and dust for giant planets during formation \citep{muzerolle2003,Dacus_2021,Muzerolle_2005}. For another unit of comparison, 10$^{-10}$ M$_\odot$/yr is approximately equivalent to 1.7 $\times 10^{5} M_\textrm{halley}/\textrm{week}$ where $M_\textrm{halley}$ is the mass of Halley's comet \citep{Cevolani1987} or also equivalent to 0.1 $M_\textrm{Jup}/\textrm{Myr}$.

\section{Results}\label{results}

The resulting model is defined by either six or eight critical parameters depending on whether or not micrometeroid dust is included. The first six are the effective temperature of the grid $T_\textrm{grid}$, surface gravity $g$, vertical eddy diffusion coefficient $K_\textrm{zz}$, condensate sedimentation efficiency $f_\textrm{sed}$, cloud hole fraction $f_\textrm{hole}$, and planetary radius $R_\textrm{planet}$. If the micrometeroid dust is considered, then the dust size distribution power law spectral index $\alpha$ and proportionality constant $N_0$ are included as well. Additionally, all of the models we consider in this paper have a fixed atmospheric metallicity and C/O ratio equivalent to the solar abundances. 

The atmospheric hole fraction is based on the patchy cloud model of \citep{Marley_2010}, and represents the relative weighting in a superposition of two distinct radiative transfer calculations, one with and the other without clouds, i.e. where the OPD, $w_0$, and $g_0$ are zero everywhere.
\begin{equation}
    F_{\lambda,\textrm{patchy}} = f_\textrm{hole} F_{\lambda,\textrm{cloudfree}} + (1 - f_\textrm{hole}) F_{\lambda,\textrm{cloudy}}
\end{equation}

We compute model spectra over a large discrete grid of parameters for both the equilibrium and disequilibrium chemistry models. The specific parameter grid our calculation are run on is
\begin{eqnarray*}
    T_\textrm{grid} &\in& [500, 600, ..., 1200, 1300] \textrm{ K}\\
    g &\in& [100, 178, 316, 562, 1000] \textrm{ m}/\textrm{s}^2 \\
    \alpha &\in& [-5, -4.75, ..., -2.25, -2] \\
    N_0 &\in& 10^{[11, 11.5, ..., 14.5, 15]} \textrm{ }\frac{\textrm{particles}}{\textrm{s} \cdot \textrm{m}^3} \\
    f_\textrm{sed} &\in& [1, 2, ..., 9, 10] \\
    K_\textrm{zz} &\in& 10^{[3, 4, ..., 9, 10]} \textrm{ cm}^2/\textrm{s }\\
    f_\textrm{hole} &\in& [.05, .1, ..., .95, 1.0] \\
    R_\textrm{planet} &\in& 10^{[-0.5, -0.45, ..., 0.45, 0.5]} R_\textrm{Jup},
\end{eqnarray*}
which are a total of 9 grid temperatures, 5 gravities, 13 dust power law indices, 9 dust power law coefficients, 10 sedimentation efficiencies, 8 vertical eddy diffusion coefficients, 21 cloud hole fractions, 21 planet radii, for a grand total of 374,673,600 unique atmospheric models.

Each model spectrum is compared to the observations of 51 Eridani b and the resulting goodness of fit metric $\chi^2$ is calculated at every point (Equation \ref{eq:chi2}). From the $\chi^2$ we infer the relative likelihood of the model at every point on the discrete simulation grid
\begin{equation}
\mathcal{L}(T_\textrm{grid},g,\alpha,N_0,K_\textrm{zz},f_\textrm{sed},f_\textrm{hole},R_\textrm{planet}) \propto e^{-\chi^2/2}.    
\end{equation}

\begin{figure}[htbp]
    \centering
    \includegraphics[width=\textwidth]{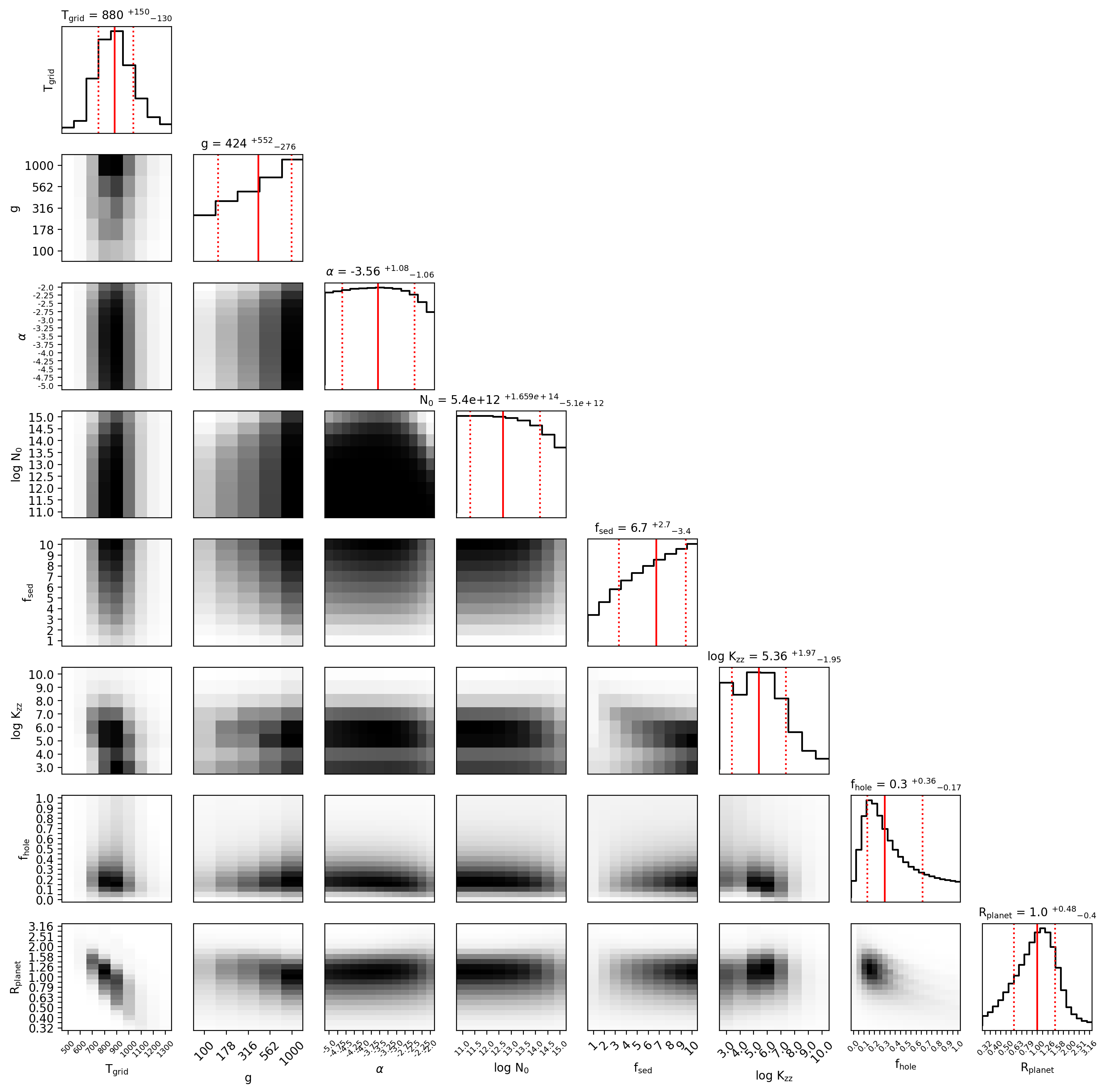}
    \caption{Discrete approximation to posterior probability distributions for disequilibrium atmospheric model parameters of 51 Eridani b. Parameter-pair covariance surfaces use a black-hot colormap, while single parameter posteriors are scaled such that the y-axis begins at likelihood $\mathcal{L} = 0$. The 16$^{\textrm{th}}$, 50$^{\textrm{th}}$, and 84$^{\textrm{th}}$ percentiles values are shown in red, with corresponding values printed in the title.}
    \label{fig:triangle}
\end{figure}

\begin{figure}[htbp]
    \centering
    \includegraphics[width=.55\textwidth]{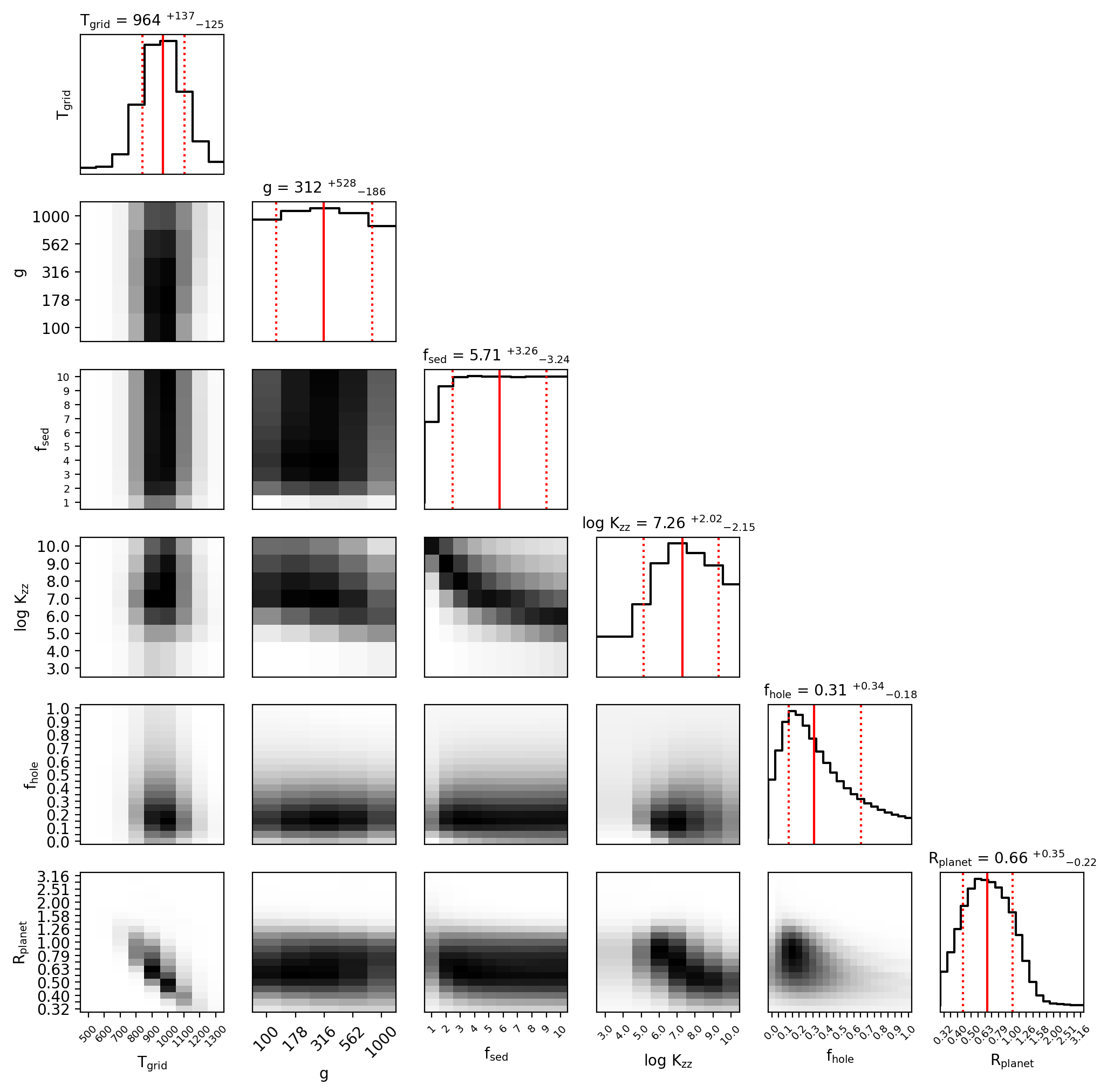}
    \caption{Equilibrium chemistry and dust free atmospheric model posterior parameter distribution triangle plot.}
    \label{fig:tri_e0}
\end{figure}

This represents an eight dimensional discrete approximation to the likelihood landscape of atmospheric parameters under the various modelling assumptions which we have made. In order to visualize this surface, we compute single-parameter and parameter-pair marginal probability distributions with exclusive summation, for example
\begin{equation}
    \mathcal{L}(T_\textrm{grid},g) \propto \sum_{K_\textrm{zz}} \sum_{f_\textrm{sed}} \sum_{f_\textrm{hole}} \sum_{\alpha} \sum_{N_0} \sum_{R_\textrm{planet}} \Big[ \mathcal{L}(T_\textrm{grid},g,K_\textrm{zz},f_\textrm{sed},f_\textrm{hole},\alpha,N_0,R_\textrm{planet}) \Big]
\end{equation}
is the inferred posterior probability distribution between grid temperature and surface gravity, allowing for visualization of possible covariance between those two parameters, if any exists in the grid search. Likewise, other parameter-pair marginal distributions simply sum the contributions from the excluded parameter axes, while single parameter distributions are marginalized over all other parameters
\begin{equation}
    \mathcal{L}(T_\textrm{grid}) \propto \sum_{\neg T_\textrm{grid}} [\mathcal{L}].
\end{equation}
These visualizations for the likelihood landscape are plotted in standard triangle format for the disequilibrium dusty atmosphere model in Figure (\ref{fig:triangle}). Additional triangle plots for equilibrium chemistry and dust free models are attached in the Appendix (Figures \ref{fig:tri_ed}, \ref{fig:tri_d0}, and \ref{fig:tri_e0}).
In order for these curves to be interpreted properly as probabilities, there is only one sensible normalization for their amplitude. The integral over the entire parameter plane for the probability density should be equal to one, however, it is not entirely clear how to calculate that integral from the inferences at the discrete grid points alone. For some of the parameters, such as $T_\textrm{grid}$, $K_\textrm{zz}$, $f_\textrm{hole}$, and $R_\textrm{planet}$, the range of values captured by the grid seems to capture the bulk of the posterior probability density, while for some parameters such as $g$, $\alpha$, $N_0$, and $f_\textrm{sed}$, it would be unwise to claim the same. The choice of a discrete grid on which to calculate these models is essential for the tractability of the problem, but also imposes a biased prior probability on the inference. Parameter values outside of the grid are functionally impossible and so the prior can be thought of as a top hat over the grid. This trade off between between tractability and completeness has no apparent resolution, and so the inferred median parameters may not accurately represent the true median over all possible parameters, but instead the median on top of the biased grid. For parameters where the majority of the posterior probability is captured, these estimates are more reliable than those which are not fully captured.

Regardless of this nuance regarding the interpretation of the median values in this landscape, it is still interesting to investigate the relationships between the parameters and their influence on the resulting model spectra, and therefore the constraints the data can place on the model parameters. One of the most critical covariance surfaces to investigate is that between $T_\textrm{grid}$ and $R_\textrm{planet}$. These parameters are highly covariant because they both strongly control the total flux emitted by the planet, and so the resulting inference constrains the pair to roughly an ellipse along a line of constant luminosity. However, comparing the centroid of this ellipse in Figure (\ref{fig:triangle}) to Figure (\ref{fig:tri_ed}) which shows the same inferences except for using the equilibrium chemistry  models, an important distinction can be made. The equilibrium chemistry models are constrained by the data to have a higher median value of $T_\textrm{grid}$ $\sim$ 970 K and lower median value of $R_\textrm{planet}$ $\sim$ 0.7 R$_\textrm{Jup}$ compared to the disequilibrium grid, which instead prefers $T_\textrm{grid}$ $\sim$ 880 K and $R_\textrm{planet}$ $\sim$ 1.0 R$_\textrm{Jup}$. This is likely tied to the fundamentally different spectral shapes which results at long wavelengths due to the presence of high altitude carbon monoxide.

Regarding the micrometeroid dust parameters, the collisional equilibrium power law index $\alpha = -3.5$ is generally the highest likelihood value, but overall the posterior is driven by the bounds of the grid parameter space. Similarly, the proportionality constant $N_0$ is not tightly constrained. Values of $\log N_0 \lesssim$ 13 have almost no apparent impact on the resulting shape of the optical spectra, and so the likelihood is roughly flat below this value, while for larger $N_0$ the dust infall is so significant that the resulting spectra are substantially different, and to some extent, excluded by the data. However, the majority of these micrometeroid parameter correspond to enormous mass rates of infalling material, see Figure (\ref{fig:mass_infall}) in the Appendix for more details. At least in the range of wavelengths currently observed with data, the micrometeroid dust only has a significant influence on the colors when the mass rates are unreasonably large.

Two of the parameters which are not well constrained are the surface gravity $g$ and the sedimentation efficiency $f_\textrm{sed}$. These parameters both influence the vertical distribution of cloud opacity sources. In the micrometeroid dust model, higher gravity implies greater terminal velocity (Equation \ref{eq:vt}) and thus greater vertical filtering of dust particles with different radii. In the cloud condensate model \citep{Ackerman_2001} sedimentation efficiency (referred to in the original paper as $f_\textrm{rain}$) controls the vertical distribution of condensate clouds, as well as particle size distributions along with the vertical eddy diffusion coefficient $K_\textrm{zz}$. The posteriors suggest that models with high gravity and $f_\textrm{sed}$ are preferable, which lessens the importance of the cloud opacity in comparison to the gas opacity as the condensates are concentrated at lower altitudes. Additionally, considering the same posteriors but in Figure (\ref{fig:tri_d0}) suggests the micrometeroid opacity is not responsible for compensating the cloud opacity in this fashion, as the effect is still present when the assumed purely scattering dust is absent. Additionally, surface gravity's influence on low resolution spectra is a very minor effect and this makes $g$ a notoriously difficult parameter to constrain.

Two of the parameters which are somewhat well constrained are the vertical eddy diffusion coefficient $K_\textrm{zz}$ and the cloud hole fraction $f_\textrm{hole}$. For the eddy diffusion coefficient the highest likelihood values are sensible, around $\sim$10$^{5.5}$ cm$^2$/s or $\sim$30 m$^2$/s, which makes it comparable to models for eddy diffusion profiles for various solar system planets \citep{Zhang_2018}, whose values range from 10$^{-1}$ to 10$^4$ m$^2$/s. Generically, a moderate hole fraction around one quarter are the highest likelihood models, although the median is skewed a bit higher due to the grid extending all the way up to $f_\textrm{hole} = 1$ which would be a completely cloudless model. These moderate hole fractions are responsible for the ``peaks" of flux at the center of each of J, H, and K bands. Models with $f_\textrm{hole} = 0$ are shown in Figure (\ref{fig:dust}) and these models generally have ``flattened" peaks. In these models the brightest regions of high flux are muted due to the presence of clouds blocking photons from the deepest, hottest layers. However, with a nonzero hole fraction photons from the deepest, hottest layers can pass through resulting in narrow wavelength regions with high flux which appear brighter, thus resulting in ``sharper peaks" in the SED. This may in part explain the unusual shape of the planet radius versus hole fraction covariance $\mathcal{L}(f_\textrm{hole},R_\textrm{planet})$, where there is a roughly three-pronged comet-shaped tail which correspond to the three spectral bands of data in J, H, and K.

\begin{figure}[htbp]
    \centering
    \includegraphics[width=\textwidth]{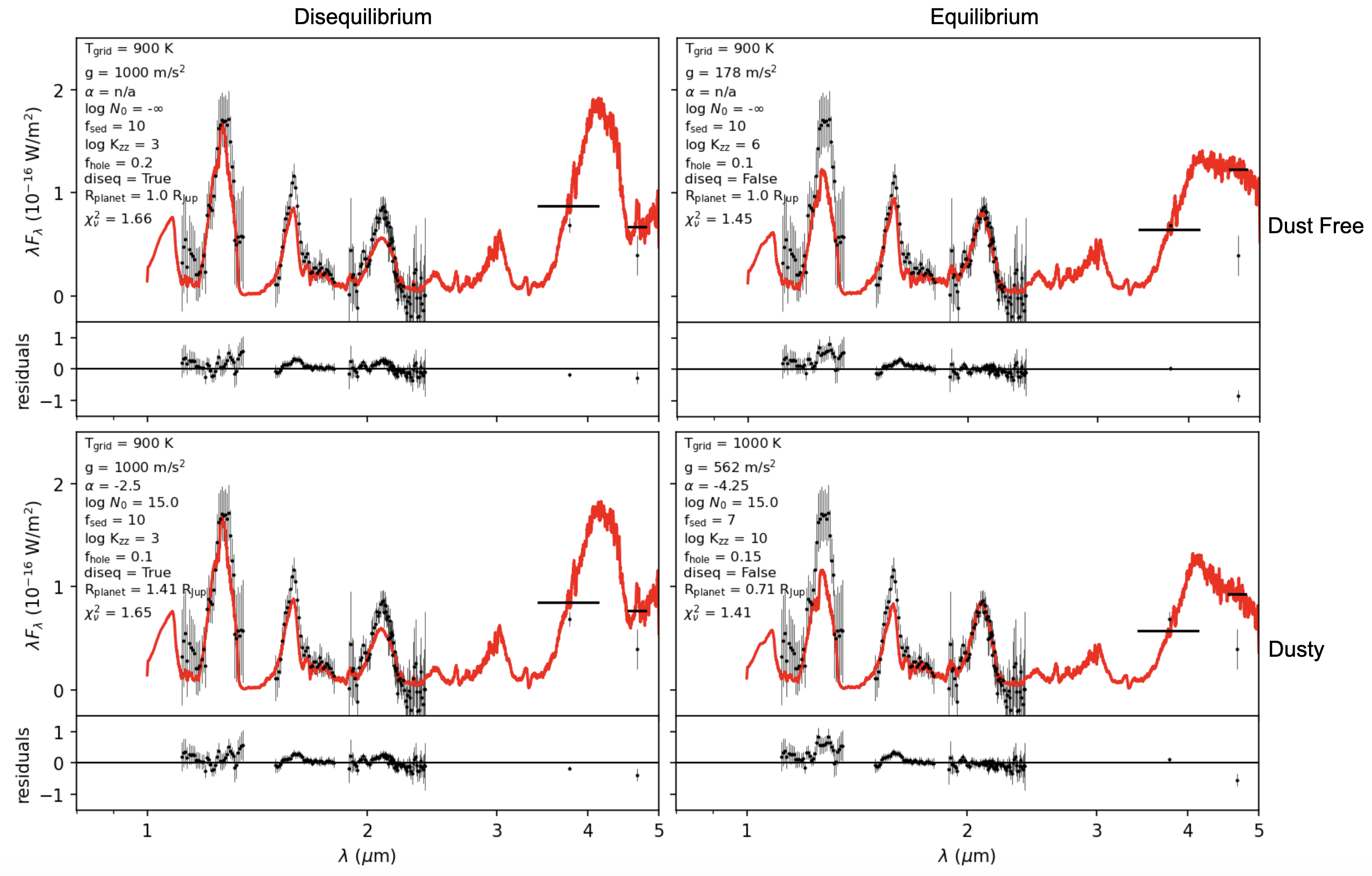}
    \caption{Best fit model spectra from each unique model class. Although the equilibrium chemistry models are optimal from the perspective of $\chi^2_\nu$, they struggle to fit the M-band photometric point. This single point does not have a large influence on the final $\chi^2_\nu$ compared to the spectroscopic bands in JHK which have a much greater information content. The dusty models generically do not perform better than the dust-free models.}
    \label{fig:spectra}
\end{figure}

While the entire landscape of possibilities proves useful to investigate, the best fit spectra from each class of model (equilibrium / disequilibrium, dusty / dust free) according to the reduced $\chi^2_\nu$ are shown for comparison in a grid layout in Figure (\ref{fig:spectra}).

Examining this figure, it is clear how the difference between the equilibrium and disequilibrium chemistry influences the long wavelength shape of the spectra. The existence of the higher altitude CO in the disequilibrium model accentuates the spectral peak around 4 $\mu$m, which is more like a cliff or plateau in the equilibrium chemistry model. This difference in spectral shape helps to consistently fit the long wavelength photometry from the Keck observations which the equilibrium chemistry struggles with. While the equilibrium models actually have lower $\chi^2_\nu$ values due to more accurate fits in K band, they underestimate the flux in J band, which is not significantly punished due to the spectral covariance of the data points, including the bright ``shoulder" around 1.1 $\mu$m which may contain residual speckle noise.

It is not clear that including the micrometeroid dust provides any significant benefit to the best fit model spectra. While the median value for the dust power law index matches the collisional cascade, both of the best fit dusty models seem to prefer unphysical power laws, with either steep or shallower slopes depending on which chemistry is being considered, which suggests the goodness of fit is actually pathological. The resulting $\chi^2_\nu$'s vary by only a few percent between the dusty and dust free cases, suggesting these degrees of freedom are not critical for fitting the spectra. And most importantly, the dusty equilibrium model has a mass rate around 10$^{-9}$ M$_\odot$/yr, while the dusty disequilibrium model has a mass rate around 10$^{-8.5}$ M$_\odot$/yr, both of which are far too large to be physically sensible.

In order to facilitate a model comparison between this work and previous attempts to model the spectra of 51 Eridani b, Table (\ref{tab:literature_comparison}) was created. While many of the individual models make unique assumptions about which parameters are included, especially in regard to the cloud parameters and composition of the condensate particles, the comparison of effective temperature, radius, gravity, and luminosity are useful to get a broad overview of the model landscape which has already been tried. In particular, regardless of cloud parameters, all of the models agree well about the planet's bolometric luminosity. It is also clear that all of these are subject to a tradeoff between radius and effective-temperature in regards to this constraint.

\begin{table}[H]
    \centering
    \begin{tabular}{c c c c c c c c c c c c}
        Eq. & Cloud & T$_\textrm{eff}$ & $\log(g)$ & $\log(K_\textrm{zz})$ & & & & R$_\textrm{planet}$ & $\log(L)$ & \\
        Chem? & Model & (K) & $(\textrm{m}/\textrm{s}^2)$ & $(\textrm{cm}^2/\textrm{s})$ & f$_\textrm{sed}$ & f$_\textrm{hole}$ & [M/H] & (R$_\textrm{Jup}$) & $(L_\odot)$ & Reference \\
        \hline
        Yes & --- & 750 & 3.5 & --- & --- & --- & 0 & 0.76 & -5.8 & \cite{macintosh2015} \\
        No & Partly-Cloudy & 700 & 1.5 & 8 & --- & 0.5 & 0 & 1 & -5.6 & \cite{macintosh2015} \\
        Yes & Iron-Silicates & 900 & 1.25 & --- & 2 & 0 & 0 & 0.57 & -5.83 & \cite{Rajan_2017} \\
        Yes & Salt-Sulfide & 725 & 2.5 & --- & 2 & --- & 0 & 0.94 & -5.93 & \cite{Rajan_2017} \\
        Yes & Salt-Sulfide & 775 & 3 & --- & 2 & --- & 0.5 & 0.72 & -5.75 & \cite{Rajan_2017} \\
        Yes & --- & 900 & 1.5 & --- & --- & --- & 0 & 0.52 & -5.77 & \cite{Rajan_2017} \\
        Yes & --- & 850 & 1.5 & --- & --- & --- & 0.5 & 0.60 & -5.75 & \cite{Rajan_2017} \\
        Yes & Cloudy & 760 & 2.26 & 7.5 & 1.26 & --- & --- & 1.11 & -5.41 & \cite{Samland2017} \\
        No* & --- & 769 & 2.26 & --- & --- & --- & -0.26 & 1.09 & -5.40 &\cite{Whiteford2023} \\
        Yes & Cloudy/Dusty & 777 & 2.75 & 10 & 7 & 0.15 & 0 & 0.71 & -5.78 & This work \\
        Yes & Cloudy/Dust-Free & 674 & 2.25 & 6 & 10 & 0.1 & 0 & 1.0 & -5.73 &This work \\
        No & Cloudy/Dusty & 575 & 3 & 3 & 10 & 0.1 & 0 & 1.41 & -5.70 & This work \\
        No & Cloudy/Dust-Free & 681 & 3 & 3 & 10 & 0.2 & 0 & 1.0 & -5.71 & This work
    \end{tabular}
    \caption{Model comparison between \citep{macintosh2015,Rajan_2017,Samland2017,Whiteford2023} and this work. Cells populated with em-dash indicate no information was available or that parameter was unused in the particular model. *The model in \citep{Whiteford2023} uses a retrieval analysis where the chemistry is arbitrary, while we populate this table with their analysis on GPI data from their Table 5.  Despite the heterogeneity of modelling choices, most of the models agree well regarding the luminosity of the object.}
    \label{tab:literature_comparison}
\end{table}

\section{Discussion}\label{discussion}

In this paper, we investigated the effects of disequilibrium chemistry and micrometeroid dust on the resulting near-infrared spectra of 51 Eridani b, and compared these models to data taken with the Gemini Planet Imager and Keck/NIRC2. We showed how vertical mixing in the atmosphere pushes carbon monoxide abundances out of equilibrium, resulting in a strong absorption feature around 4.5 $\mu$m, and investigated how differences in micrometeroid dust parameters effectively smoothly redden the entire spectra over the range of wavelengths between 1 and 5 $\mu$m. We computed an extremely large grid of atmospheric models which are compared to the data, and the resulting likelihood of each model is evaluated in order to generate posterior probability densities over the model parameter space. We find that disequilibrium chemistry is useful to explain the mid-infrared colors of 51 Eridani b, especially the M band photometric point around 4.5 $\mu$m, but that micrometeoroid dust doesn't provide any additional useful degrees of freedom to explain the data. While the collisional cascade index $\alpha = -3.5$  was the median inferred value, the best fit dusty models preferred unphysical power law distributions and enormous mass infall rates to have a significant optical effect in these wavelengths. While micrometeroid dust is not necessary to explain the infrared colors of 51 Eridani b, other planets in different circumplanetary environments could still show evidence of micrometeroid dust, especially at longer wavelengths such as around 10 $\mu$m where SiO$_2$ has a significant absorption feature (Figure \ref{fig:n}).

The best fitting, dust free, disequilibrium chemistry model has the following parameters $T_\textrm{grid}$ = 900 K, $g$ = 1000 m/s$^2$, $f_\textrm{sed}$ = 10, $K_\textrm{zz}$ = 10$^3$ cm$^2$/s, $f_\textrm{hole}$ = 0.2, and $R_\textrm{planet}$ = 1.0 R$_\textrm{Jup}$. However, an important clarification regarding the interpretation of the grid temperature is that it corresponds to the effective temperature for a cloudless atmosphere with only gas opacity, before cloud profiles are post-processed during the radiative transfer. After these are calculated the resulting spectral energy distribution can be integrated over all wavelengths to find the bolometric luminosity $L = \int F_\lambda \textrm{d}\lambda = 4\pi R_\textrm{planet}^2 \sigma T_\textrm{eff}^4$, and therefore the true effective temperature including the effects of the post-processed clouds. Under this interpretation, we find the best fit dust free disequilibrium model has effective temperature 681 K and bolometric luminosity $\log_{10} \frac{L_\textrm{planet}}{L_\odot}$ = -5.71, which agree well with previously reported results from \citep{Rajan_2017} which report T$_\textrm{eff}$ between 605 and 737 K and luminosity between -5.83 to -5.93. The moderate increase in luminosity is likely a due to the larger intensity peak in the mid infrared as a result of the disequilibrium chemistry abundance of carbon monoxide.

Our modeling framework is a first step toward incorporating micrometeroid dust into models, but there are many assumptions that can still be improved. The vertical eddy diffusion coefficient $K_\textrm{zz}$ is assumed constant with pressure altitude and across all chemical species, whereas a better model will account for differences in $K_\textrm{zz}$ at different altitudes \citep{mukherjee2022b} and for different chemical species \citep{Zhang_2018}, which result from interactions between vertical transport and horizontal non-uniformities in the atmosphere. Likewise, the sedimentation efficiency $f_\textrm{sed}$ as a single parameter cannot capture all of the complexity associated with the unique microphysics of cloud aerosols of different species. $f_\textrm{sed}$ may even vary within different condensate populations of the same species, for example both hail and snowflakes are composed of water ice but form in unique conditions. 

The superposition assumption of a cloudy and cloudless atmosphere with a relative weighting of the hole fraction $f_\textrm{hole}$ is not entirely-self consistent due to radiative feedback which alters the temperature pressure profiles in the two cases. In general, the superposition assumption which we use to combine the cloud condensate opacity model with the micrometeroid opacity model assumes these populations of material do not interact, while significant material entering the planet's atmosphere from space could significantly perturb the chemical abundances and radiative profiles of the model, especially in a time-dependent fashion. If the micrometeroid populations are to be considered realistic, they should vary with space and time in some manner, although its not entirely clear how one would estimate this variability, simulations of rocky material in near-planetary orbits could be investigated to better constrain the variability properties of the infalling dust. The dust model itself could be improved beyond purely spherical Mie-scattering grains of silica, accounting for different modes of agglomeration, different compositions and therefore optical properties, along with the time and space variability, and chemical and radiative feedback. Many of these imperfect assumptions were made with tractability in mind, as a completely self-consistent atmosphere model is well beyond any reasonable scope. Even if one could solve the fully three dimensional coupled, hydrodynamic-radiative-chemical equations, there would still ultimately remain questions about how to parameterize sub-grid-scale dissipative effects, among other limitations which remain in any discretized model. Ultimately, models of extrasolar planet atmospheres are the interpretation machinery which connect observable colors to planetary parameters such as the thermal profile or its chemical abundances. Further improvements to these models, alongside additional empirical data such as M-band spectroscopy, will be necessary to gain deeper insights into the turbulent dynamics inside these natural laboratories. 

\section{Appendix: physical and optical properties of micrometeroid dust}

The interactions between planets and the interplanetary environment are numerous and complex. In the beginning of their life, protoplanets grow through gravitational accretion of disk material 
\citep{Valletta2021}, with an estimated radius of influence which depends on the planet's composition among other factors. Further evidence for gravitational capture remains long after planet formation as irregular satellites with large, eccentric, and inclined orbits, which all of the giant planets in the solar system possess \citep{Jewitt2007}. This is in opposition to the formation mode of circumplanetary accretion which instead produces satellites with circular orbits with relatively low inclinations. It is possible that satellites exist within the magnetosphere \citep{Mendis1974} of their host planets, resulting in fascinating phenomena such as the Io-Jupiter decametric radiation. 

Furthermore, the electromagnetic dynamics of small dust grains in the interplanetary environment are significant. Charged dust dynamics can result in levitation, rapid transport, energization, ejection, capture, and the formation of new planetary rings \citep{Horanyi1996}.  Magnetospheric effects may enhance up to a factor of four the micrometeroid flux of particles colliding with Jupiter around 0.5 $\mu$m, while shielding a planet from impactors around 0.1 $\mu$m \citep{Colwell1996}.

The possibility of a near-Earth belt of dust has been investigated thoroughly including the effects of gravitational focusing, capture, radiation pressure, electromagnetic forces, hydrodynamic atmospheric drag, and enhancement from lunar ejecta
\citep{Colombo1966_1, Colombo1966_2, Colombo1966_3, Colombo1966_4}. The study found no convincing mechanism to explain the observed factor of 10$^4$ enhancement over the interplanetary background levels. However, the vaporization of lunar regolith due to impacts from micrometeroids \citep{Pokorny2019} is a suitable explanation for the existence of a rareified lunar exosphere, although the impactor flux is much smaller on the moon than compared to the Earth due to its smaller cross section and lower gravitational focusing factor.

\subsection{Dust size and mass distribution}

Regardless of existing puzzling observations and uncertainties in modeling, it is clear that the interplanetary environment is not pristine empty space. The micrometeroid dust environment is thought to emerge from a collisional process of asteroidal debris which is in a steady state equilibrium between agglomeration or inelastic collision and fragmentation or shattering \citep{Dohnanyi1969}. The end result is a distribution of particles with a characteristic power law number density profile which can either be a function of particle radius or mass

\begin{equation}
    N(r_p) = A r_p^{\alpha} \textrm{d}r_p,
\end{equation}
or 
\begin{equation}
    N(m_p) = B m_p^{\beta} \textrm{d}m_p.
\end{equation}

Reported values in the literature vary mildly, in \citep{Dohnanyi1969} $\beta = -1.837$, while in \citep{Gaspar2012} $\alpha = -3.65$ and $\beta = -1.88$, and in \citep{Pan_2012} which accounts for self-consistent particle velocities $\alpha$ can go to $-4$, while for large bodies which are held together with self gravity the power law can be modified from $-2.88 > \alpha > -3.14$ to $\alpha = -3.26$. 

\begin{figure}[htbp]
    \centering
    \includegraphics{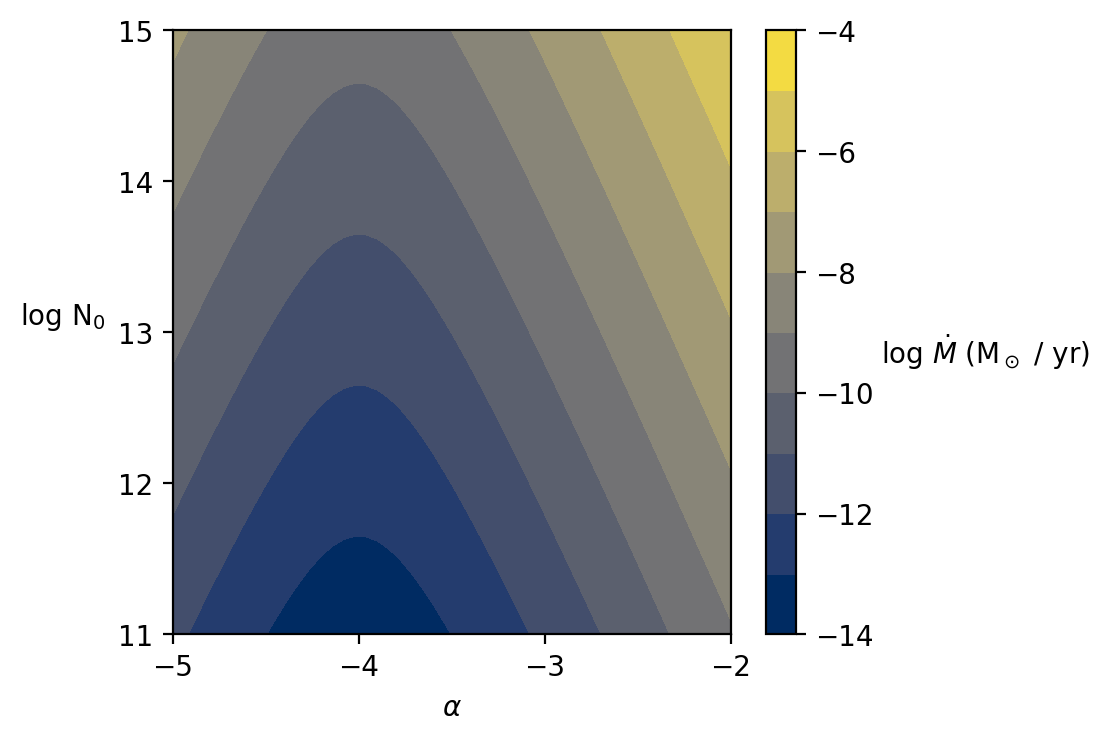}
    \caption{The total in-falling dust mass as a function of dust power law model parameters.}
    \label{fig:mass_infall}
\end{figure}

For the two formulations, the relationship between $A, B$ and $\alpha, \beta$ are simple. Assuming a spherical particle of constant density, $m(r_p) = \rho_p \frac{4}{3}\pi r_p^3$, then d$m = \rho_p 4\pi r_p^2$d$r_p$, and
\begin{equation}
    N(m(r_p)) = B \big(\rho_p \frac{4}{3} \pi r_p^3\big)^{\beta} \rho_p 4\pi r_p^2 \textrm{d}r_p
\end{equation}
then
\begin{equation}
    N(r_p) = B \frac{(4\pi\rho)^{\beta + 1}}{3^{\beta}} r_p^{3\beta + 2} \textrm{d}r_p
\end{equation}
and so $A = B \frac{(4\pi)\rho^{\beta + 1}}{3^{\beta}}$ and $\alpha = 3\beta + 2$. In this paper, we use the expression for number density as a function of radius, and consider a population of in-falling dust per unit surface area of the atmosphere d$A$ per unit time d$t$ given by $N(r_p)$, and we use a reference radius of $r_0 = 1$ $\mu$m.
\begin{equation}
    \frac{\textrm{d}N(r_p)}{\textrm{d}t \textrm{ d}A} = N_0 \Big(\frac{r_p}{r_0}\Big)^\alpha \textrm{d}r_p.
\end{equation}
$N_0$ is a constant with units of s$^{-1}$ m$^{-3}$ which controls the rate of particle flux. The total number of particles falling into the atmosphere in some range of radii [$r_1$, $r_2$] per unit area per unit time is 
\begin{equation}
    \frac{N_\textrm{particles}}{\textrm{d}t\textrm{ d}A} = \int_{r_1}^{r_2} N_0 \Big(\frac{r_p}{r_0}\Big)^\alpha \textrm{d}r_p,
\end{equation}
where we consider particles in the size range between $r_1 = 10^{-9}$ m and $r_2 = 10^{-3}$ m, or between nanometer and millimeter sizes. By multiplying the number density by the individual particles masses $m_p = \rho_p \frac{4}{3} \pi r_p^3$, the total mass of the in-falling particles can be found by
\begin{equation}
    \frac{M_\textrm{particles}}{\textrm{d}t\textrm{ d}A} =\int_{r_1}^{r_2} \rho_p \frac{4}{3} \pi r_p^3 N_0 \Big(\frac{r_p}{r_0}\Big)^\alpha \textrm{d}r_p,
\end{equation}

\begin{equation}
    \dot{M} = 4\pi R_P^2 \frac{M_\textrm{particles}}{\textrm{d}t \textrm{ d} A}
\end{equation}
where $R_P$ is the radius of the planet, which is plotted in Figure (\ref{fig:mass_infall}). 

\subsection{Atmospheric Transport}

The atmosphere model grid contains temperature as a function of pressure T(P) and tables of molecular mixing ratios $R_\textrm{mix}(\chi, P)$ for various molecular species $\chi$ at each pressure altitude. The combination of these parameters will be useful for determining the fluid density of the atmosphere as well as the corresponding physical radial coordinate of the various pressure layers. By taking the mixing ratio weighted sum of the atomic mass of the various species $\chi$, one obtains the mean molecular mass of each pressure altitude.

\begin{equation}
    \langle M_\textrm{amu} \rangle(P) = \frac{\sum_\chi R_\textrm{mix}m_\textrm{amu}}{\sum_\chi R_\textrm{mix}}
\end{equation}

\begin{figure}[htbp]
    \centering
    \includegraphics[width=\textwidth]{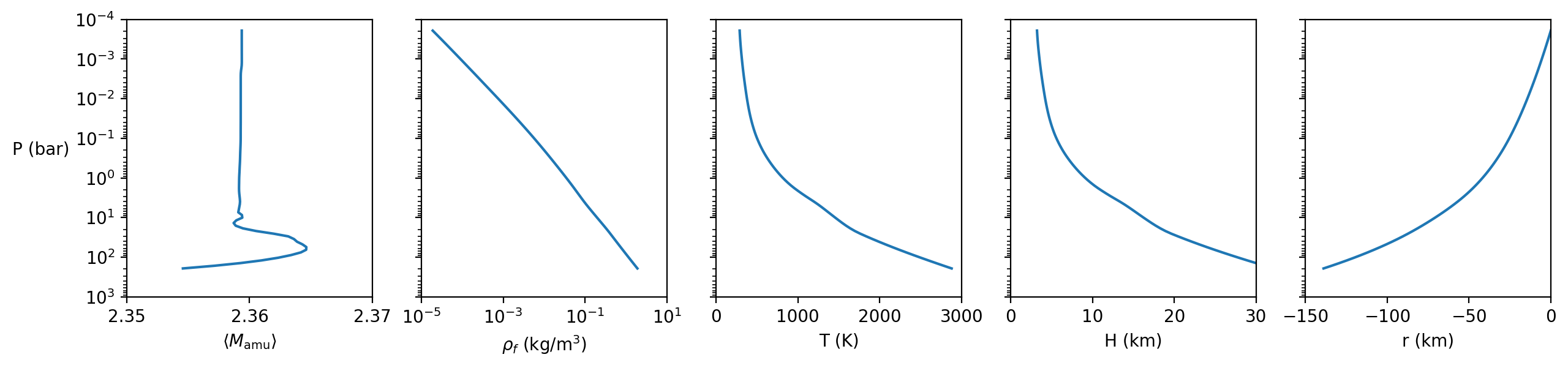}
    \caption{Atmospheric parameters as a function of pressure altitude.}
    \label{fig:pt}
\end{figure}

The mean molecular mass (amu / molecule) can be converted to a molar mass (kg / mol) by simply multiplying by the atomic mass unit $1$ $\textrm{amu} = 1.66 \times 10^{-27}$ kg and Avogadro's constant $N_A = 6.022 \times 10^{23}$ (molecules / mol), which is convenient for use with the ideal gas law
\begin{equation}
    \rho_f = \frac{P}{T} \frac{M_\textrm{molar}}{R},
\end{equation}
where $R = 8.314$ J K$^{-1}$ mol$^{-1}$ is the molar gas constant. The fluid density is useful not only to compute the terminal velocity of a spherical grain of silicate rock but also to infer the relationship between the physical radial coordinate which remains undefined and the pressure coordinate which is used in the model. Assuming the gas is in hydrostatic equilibrium \citep{marley_robinson_2015}
\begin{equation}
    \frac{\textrm{d}P}{\textrm{d}r} = -g \rho_f,
\end{equation}
then the radial coordinate can be found directly by integration
\begin{equation}
    r = \int_{P_0}^{P_1} \frac{-1}{g \rho_f} \textrm{d}P + C,
\end{equation}
where $P_0 = 10^{-4}$ bar and $P_1 = 10^2$ bar are the edges of the pressure grid and the constant $C$ is chosen so that the zero coordinate is at the top of the atmosphere. This arbitrary constant will be irrelevant as change in radial coordinate is the only relevant quantity for determining the particle size distribution as a function of pressure altitude. The mean molecular mass, fluid density, temperature, and radial coordinates are plotted alongside the atmospheric scale height $H = \frac{RT}{gM_\textrm{molar}}$ as a function of pressure altitude in Figure (\ref{fig:pt}).

Neglecting the details of gravitational capture and orbital velocity, a simple assumption is that the particles fall through the atmosphere of the planet at their terminal velocity. The terminal velocity for a particle of density $\rho_p$ and radius $r_p$ falling through a fluid of density $\rho_f$ can be found with \citep{dey2019}
\begin{equation}
    v_t = \sqrt{\frac{8gr_p}{3C_d}\Big(\frac{\rho_p}{\rho_f} - 1 \Big)}
    \label{eq:vt}
\end{equation}
where $g$ is the surface gravity, $C_d \approx 0.5$ is the drag coefficient of a spherical particle \citep{Munson2007} for flow with Reynold's number between 10$^3$ and 10$^5$, and the density of silicon dioxide is roughly $\rho_p = 2200$ kg/m$^3$ \citep{haynes2011crc}. Since the ratio of particle to fluid density $\frac{\rho_p}{\rho_f}$ can range from 10$^3$ to 10$^8$, it is reasonable to include the approximation that $\frac{\rho_p}{\rho_f} >> 1$ which is equivalent to neglecting the influence of the buoyant force due to fluid displacement. 
\begin{figure}[htbp]
    \centering
    \includegraphics[width=\textwidth]{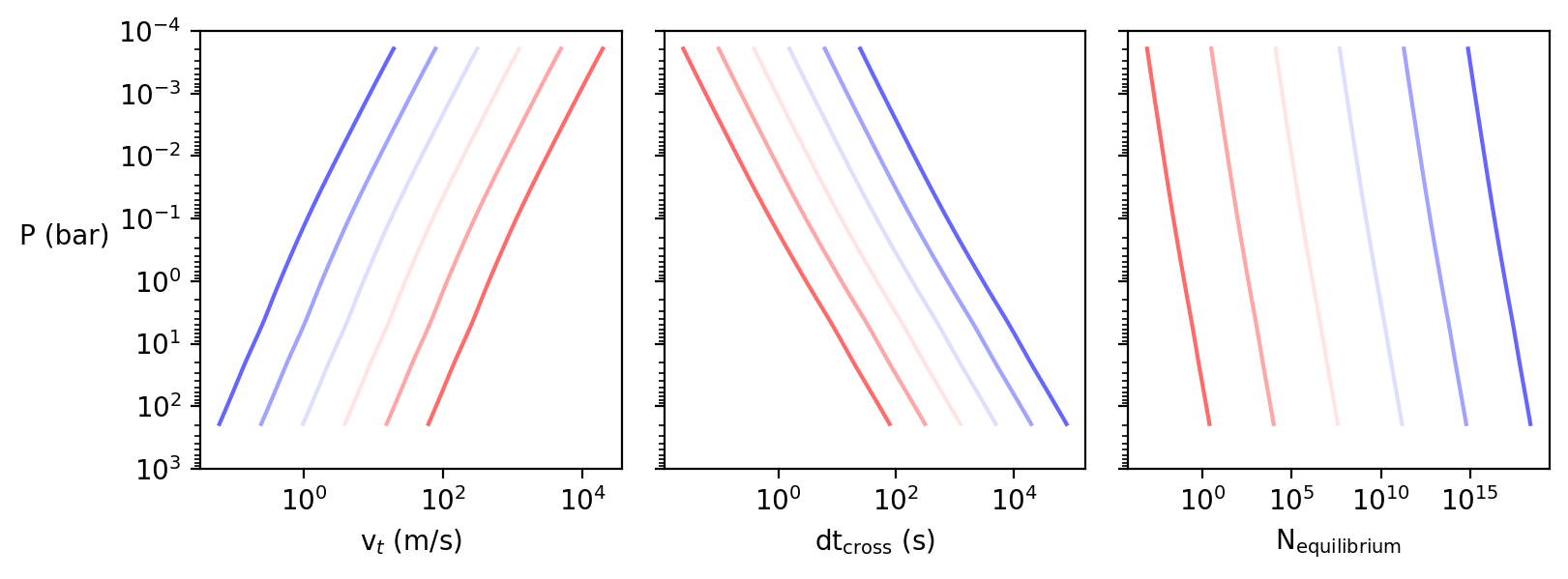}
    \caption{Dust transport values as a function of pressure altitude and particle radius for a particular atmospheric realization with grid temperature $T_\textrm{grid} = 800$ K, surface gravity $g = 316$ m/s, dust power law index $\alpha = -3.5$, and dust power law coefficient $\log N_0 = 12$. Colored curves correspond for particles between radii r$_1$ = 1 nm and r$_2$ = 1 mm with blue being the smallest particles and red for the largest with a logarithmic spacing between different lines.}
    \label{fig:terminal_velocity}
\end{figure}
With the terminal velocity $v_t$ of the particles at each radius and pressure altitude and the radial displacement of each pressure layer d$r$, one can infer the timescale for each particle to cross each pressure layer by falling through: $\textrm{d}t_\textrm{cross} = \textrm{d}r / v_t$. Combined with the initial infall rate at the top of the atmosphere $N_\textrm{particles}$ one can infer the equilibrium number of particles at each pressure altitude for every particle size. The terminal velocity $v_t$, crossing timescale d$t_\textrm{cross}$, and equilibrium number density of particles N$_\textrm{equilibrium}$ are plotted in Figure (\ref{fig:terminal_velocity}). 

\begin{equation}
    N_\textrm{equilibrium} (P,r_p) = N_\textrm{particles}(r_p) \textrm{d}t_\textrm{cross} (P,r_p)
\end{equation}

\subsection{Dust optical properties}

With the equilibrium number densities of the particles at every altitude, all that is left to infer their influence on the resulting spectra is some information about their optical properties, including the single scattering albedo, optical depth per layer, and the asymmetry factor, which are inputs to the \texttt{PICASO} radiative transfer scheme.  These are obtained via a parametric approximation to the index of refraction as a function of wavelength and the use of Bohren and Huffman's mie scattering program \texttt{bhmie} \citep{bohrenhuffman} which has been translated into python courtesy of Herbert Kaiser \citep{bhmiepython}.

\begin{figure}[htbp]
    \centering
    \includegraphics[width=.9\textwidth]{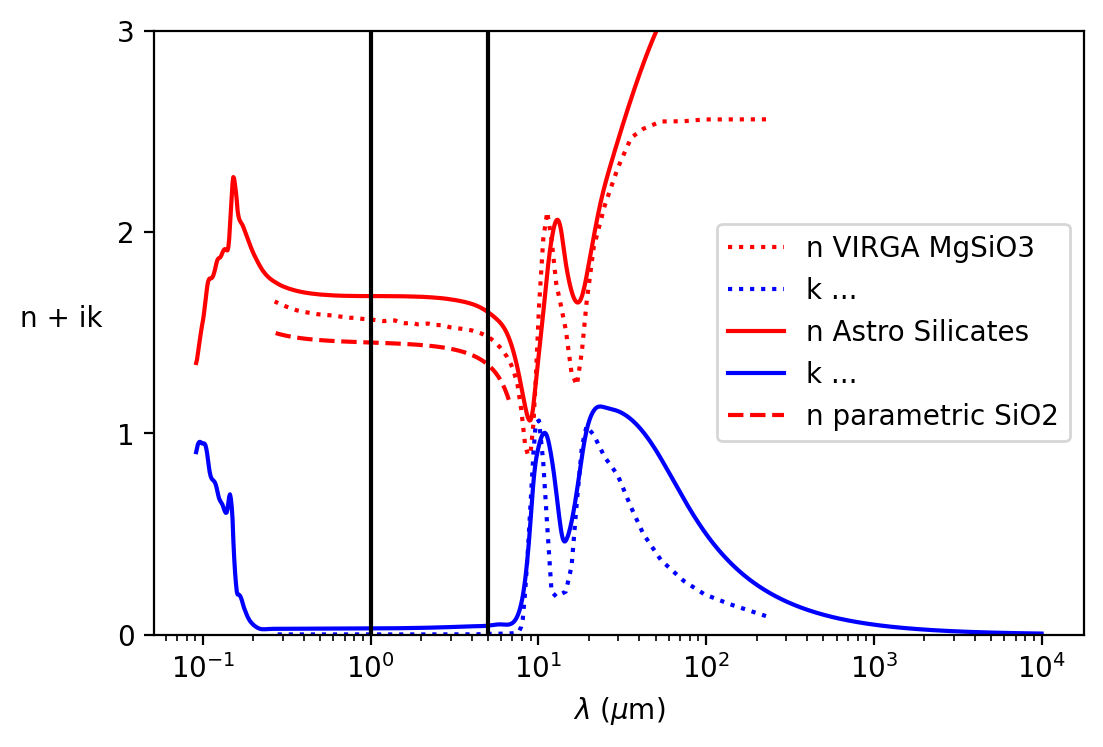}
    \caption{Index of refraction for silicates from different sources. The black vertical lines indicate the range of wavelengths important to this study.}
    \label{fig:n}
\end{figure}

The real part of the index of refraction of silicon dioxide is based on room temperature empirical measurements \citep{Malitson65} for the range of 0.21 to 3.71 $\mu$m:

\begin{equation}
    n^{2} - 1 = \frac{0.6961663 \lambda^{2}}{\lambda^{2} - (0.0684043)^{2}}
              + \frac{0.4079426 \lambda^{2}}{\lambda^{2} - (0.1162414)^{2}}
              + \frac{0.8974794 \lambda^{2}}{\lambda^{2} - (9.896161 )^{2}},
\end{equation}
while the imaginary part is assumed to equal zero. These empirical measurements agree well with later measurements in the range \citep{TAN1998} of 3 to 6.7 $\mu$m, as well with the unified model of interstellar dust \citep{li1997} in the range of wavelengths of importance for this study, between $1$ and $5$ micron. A comparison of the index of refraction for the parametric model, the interstellar dust, and MgSiO$_3$ is in Figure (\ref{fig:n}).

The \texttt{bhmie} code takes as input the complex index of refraction of the (assumed spherical) silicate particles, as well as a size parameter $x = \frac{2\pi}{\lambda} r_p$ which depends on the wavelength of light $\lambda$ under consideration and the radius of the particles, and returns values of the scattering efficiency $Q_\textrm{sca}$ and asymmetry parameter $g_\textrm{sca}$ at every $x$. 

\begin{figure}[htbp]
    \centering
    \includegraphics[width=\textwidth]{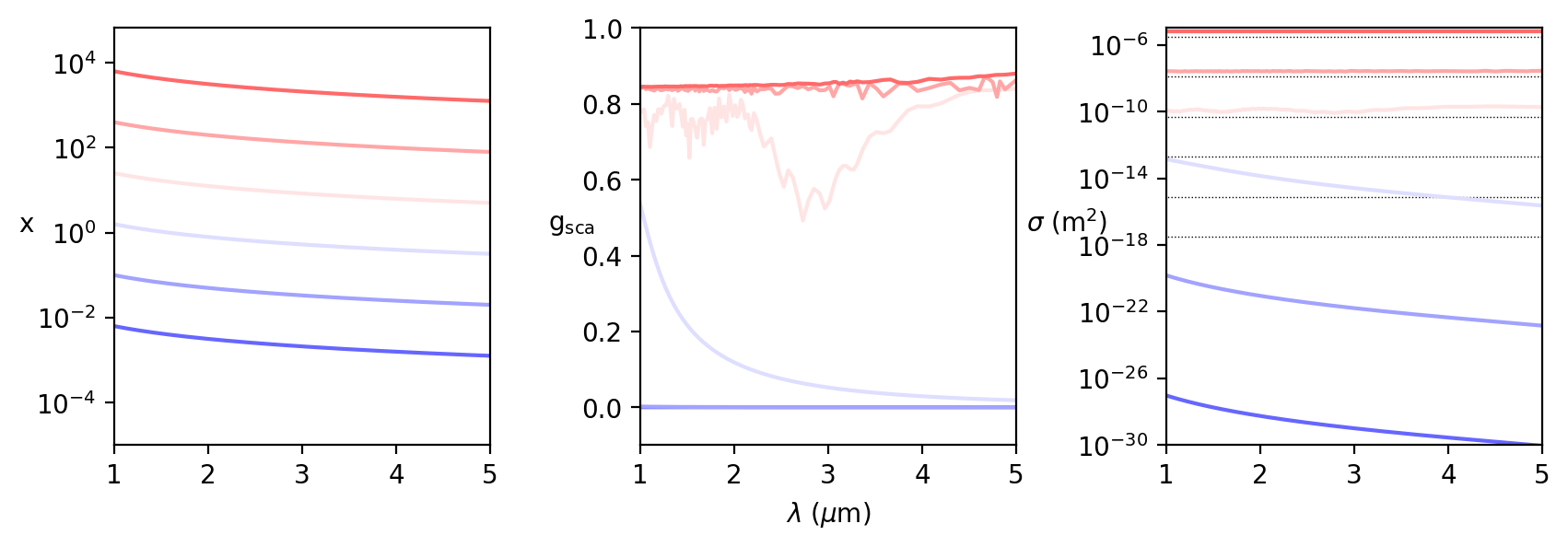}
    \caption{Scattering parameters as a function of wavelength and particle size. As in Figure (\ref{fig:terminal_velocity}), blue corresponds to nm-sized particles while red corresponds to mm-sized particles. One can see that the mm-sized particles strongly forward scatter with a scattering asymmetry parameter of $g \approx 0.8$, while the nm-sized particles scatter nearly isotropically with $g \approx 0$. In the right panel, the horizontal dashed lines correspond to the geometric cross section for an equivalently sized sphere. For particles of size with the same order of magnitude as the wavelength, the mie and geometric cross section are very similar, but the cross section drops off dramatically for the nm-sized grains and especially for longer wavelengths. This fact alongside the effect of modifying the dust power law index $\alpha$ allows the dust to modify the reddening of the spectra demonstrated in Figure (\ref{fig:dust}).}
    \label{fig:mie}
\end{figure}

The scattering efficiency is first converted into the scattering cross-section per particle $\sigma_\textrm{sca}(r_p,\lambda)$ = $Q_\textrm{sca}(r_p,\lambda) \pi r_p^2$ at every wavelength, and then to the total optical depth per layer by summing over the contributions from all particles at that altitude of all sizes under consideration.
\begin{equation}
    \textrm{OPD}(P,\lambda) = \sum_{r_p} \sigma_\textrm{sca}(r_p,\lambda) N_\textrm{equilibrium}(P,r_p)
\end{equation}
Similarly, the asymmetry parameter per layer is computed by the optical-depth-weighted sum of the asymmetry parameter for the various particle radii.
\begin{equation}
    g_0(P,\lambda) = \frac{\sum_{r_p} g_0(r_p,\lambda) \sigma_\textrm{sca}(r_p,\lambda) N_\textrm{equilibrium}(P,r_p)}{\textrm{OPD}(P,\lambda)}
\end{equation}
The single scattering albedo $w_0 (P, \lambda)$ is set to 1 everywhere for the micrometeroid dust. This is equivalent to the assumption that the imaginary part of the index of refraction is zero everywhere, that the dust is totally non-absorbing, and that all optical depth is due to scattering. The micrometeroid cloud model parameters $w_0$, OPD, and $g_0$ as a function of pressure altitude $P$ and wavelength $\lambda$ appear in the main body of the text in Figure (\ref{fig:cloud}), while the size parameter $x$, asymmetry parameter $g_\textrm{sca}$, and single particle scattering cross section are shown in Figure (\ref{fig:mie}) as a function of particle size $r_p$ and wavelength $\lambda$.

\newpage
\subsection{Additional Figures}

\begin{figure}[htbp]
    \centering
    \includegraphics[width=\textwidth]{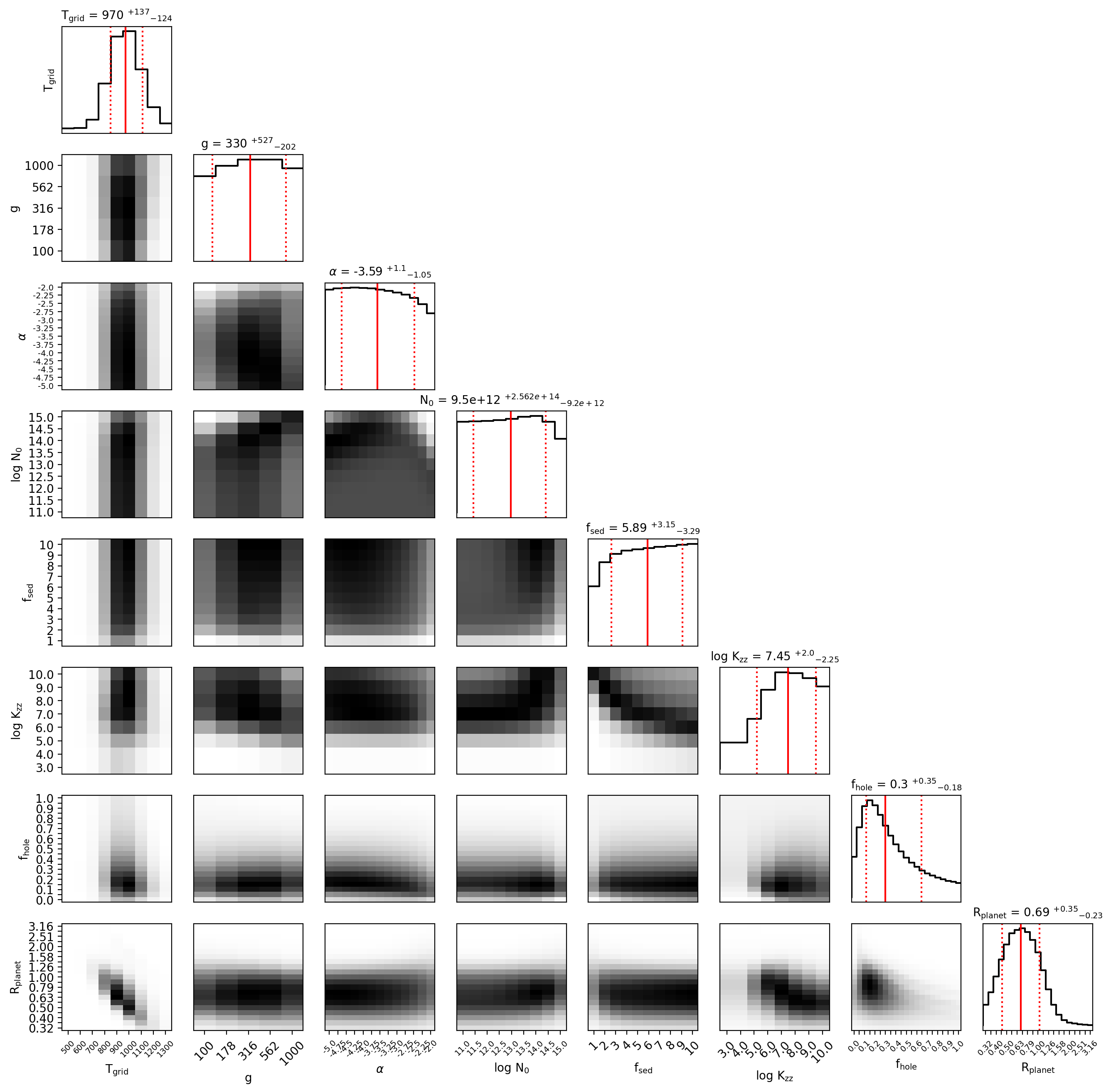}
    \caption{Equilibrium chemistry and dusty atmospheric model posterior parameter distribution triangle plot.}
    \label{fig:tri_ed}
\end{figure}
\begin{figure}[htbp]
    \centering
    \includegraphics[width=.55\textwidth]{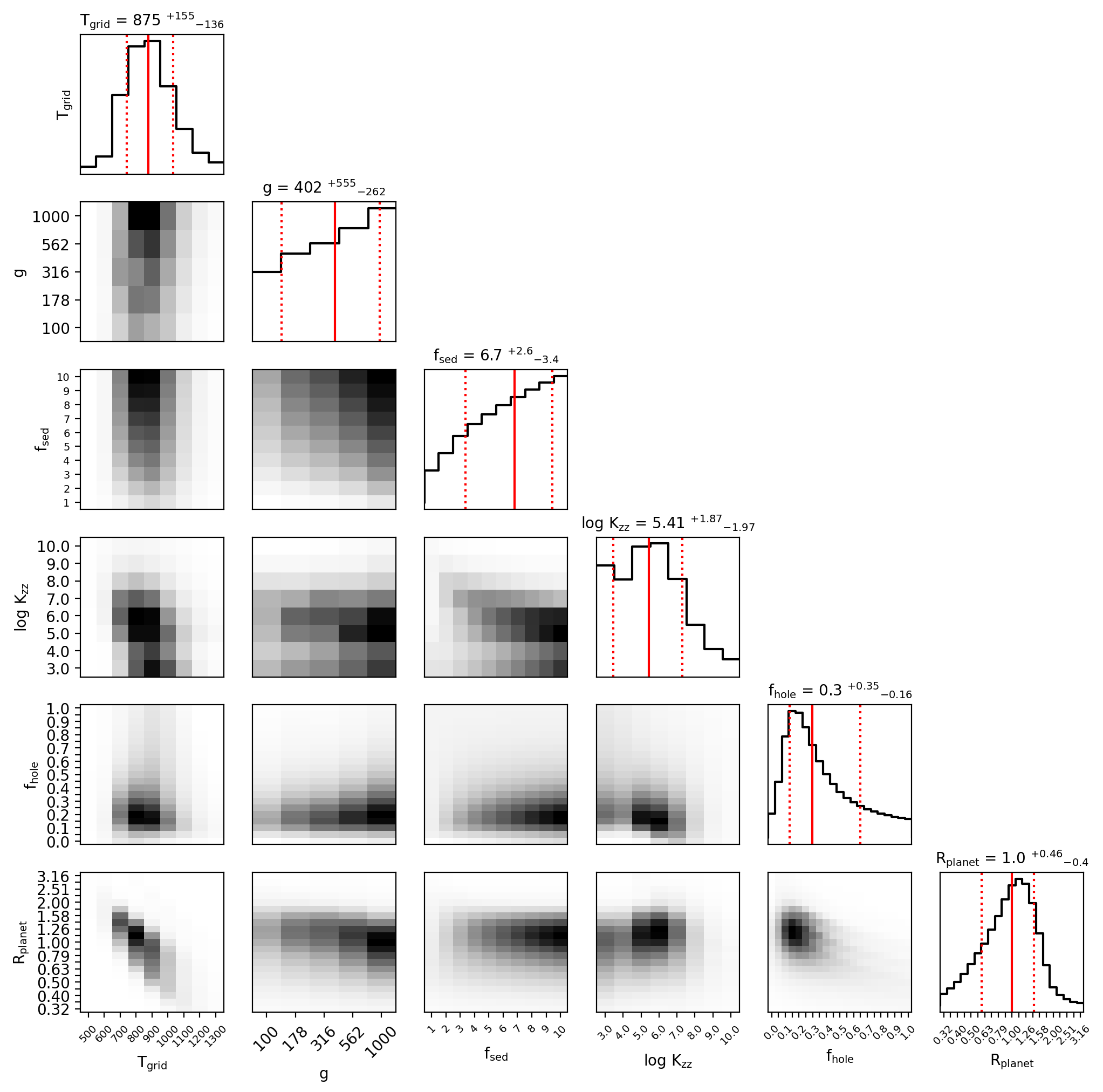}
    \caption{Disequilibrium chemistry and dust free atmospheric model posterior parameter distribution triangle plot.}
    \label{fig:tri_d0}
\end{figure}


\newpage
\bibliography{main}{}
\bibliographystyle{aasjournal}

\end{document}